# Signal Voids of Active Cardiac Implants at 3.0T CMR

Theresa Reiter[1,*], Ingo Weiss[2], Oliver M. Weber[3], Wolfgang R. Bauer[1]

[1] University Hospital Wuerzburg, Department of Internal Medicine I, Cardiology, Wuerzburg, Germany

[2] BIOTRONIK SE & Co. KG, Berlin, Germany

[3] Philips GmbH, Hamburg, Germany

* Corresponding author: Dr. Theresa Reiter, MD, University Hospital Wuerzburg, Department of Internal Medicine I, Cardiology, Oberduerbacher Strasse 6a, 97080 Wuerzburg, Germany. Reiter_t@ukw.de.

**Abstract**

Purpose

Recent technical advancements allow cardiac MRI (CMR) examinations in the presence of so-called MRI conditional active cardiac implants at 3.0T. However, the artifact burden caused by susceptibility effects remain an obstacle.

Methods

All measurements were obtained at a clinical 3.0T scanner using an in-house designed cubic phantom and optimized sequences for artifact evaluation (3D gradient echo sequence, multi-slice 2D turbo spin echo sequence). Reference sequences according to the American Society for Testing and Materials (ASTM) were additionally applied. Four representative active cardiac devices and a generic setup were analyzed regarding volume and shape of the signal void. For analysis, a threshold operation was applied to the grey value profile of each data set.

Results

The presented approach allows the evaluation of the signal void and shape even for larger implants such as ICDs. The void shape is influenced by the orientation of the B0-field and by the chosen sequence type. The distribution of ferromagnetic material within the implants also matters. The void volume depends both on the device itself, and on the sequence type. Disturbances in the B0 and B1 fields exceed the visual signal void.

Conclusions

This work presents a reproducible and highly defined approach to characterize both signal void artifacts at 3.0T and their influencing factors.

The manuscript was preprinted on Arxiv.org with the permanent identifier 2108.00437.

**Introduction**

Active cardiac implants such as pacemakers (PM) and implantable cardioverter-defibrillators (ICD) offer protection against the effects of arrhythmias and the risk of sudden cardiac death in patients with structural pathologies of the heart[1-3]. Once implanted, these implants have long been considered an absolute contraindication for the assessment of myocardial structure and function with cardiac MRI (CMR). This prevents the (re-)assessment of cardiac morphology, acute and chronic damage as well as an exact evaluation of the left and right ventricular function by CMR during the further clinical course of these patients [4,5]. The electric and ferromagnetic components of active cardiac devices interact with the strong magnetic and electromagnetic fields of the MR system and potentially cause their malfunctioning and permanent damage [6]. The most recent developments of so-called MR conditional devices have largely overcome these technical challenges under well-defined conditions [7-10]. The combination of a specific device system, device programming and a defined MR setting would allow a safely performed CMR examination. Furthermore, it has been shown in larger studies and from real-world experience, that CMR examinations with conventional devices, especially pacemakers, are feasible even though these devices are not accredited for such examinations by the manufacturers[11-13]. However, the artifact burden induced by the implant itself continues to limit the applicability of CMR scans in patients with such devices [14,15].

The presence of active cardiac implants induces complex susceptibility artifacts that appear in two ways: first, the ferromagnetic material produces a signal void in proximity to the implant; and second, the visible regions are affected by geometric distortions throughout the visible volume [16,17]. The artifact burden increases with the strength of the magnetic field, and thus poses a greater challenge for successful imaging at 3.0T. At the same time, the higher field strength offers a higher signal to noise ratio than 1.5T scanners do, and thus provides the potential to allow faster image acquisition and higher image resolution [18].

At 1.5T, approaches to minimize the artifacts during CMR imaging have been developed. Modifications of established protocols are e.g. the use of spoiled gradient echo sequences after the administration of contrast agents for CINE imaging, rather than using balanced steady state free precession (bSSFP) sequences. Broad band pulses for late enhancement imaging have been used to obtain a diagnostic image quality in the presence of active cardiac devices [19-21]. However, these modifications have not yet widely been transferred to 3.0T. Clinical feasibility studies have shown the potential of modified sequences for CMR at 3.0T but they do not systematically evaluate the quantitative extent of susceptibility artifacts of different devices [15,20].

The different image properties of the susceptibility artifacts must be analyzed separately because they require different measurement setups. On the one side, a homogeneous background for the evaluation of the signal void is required. On the other side, means to quantify the geometrical changes due to distortion must be provided. The present work focuses on the signal void and aims to establish an approach to the precise three-dimensional determination of the dimensions and the shape of signal void artifacts in phantom measurements as a first step towards image optimization at 3T CMR. For this approach, generic sequences that are the foundation of currently established clinical sequences are used. Aspects of geometric distortions are discussed in another upcoming publication.

**Materials and Methods**

Phantom

The analysis of the signal void requires a sufficiently large phantom that is capable to contain the signal void in the three spatial directions and that offers a homogenous background. A cubic phantom was designed that was filled with a homogeneous aqueous medium. The outer dimensions, as dictated in two dimensions by the 60 cm bore diameter of the scanner, measure 28x28x28 cm. In order to minimize B1-shading artifacts, while maintaining an acceptable image contrast within the water filling, a conductivity of approximately 0.5 S/m was obtained through adding 2.5 g/l NaCl. In congruency with the ASTM standard, 1g/l $Cu_2SO_4$ was added to achieve reasonable T1 times. Since no tissue types were modeled for the phantom, the relaxation times were not adjusted to physiological values. Choosing a homogeneous background, additional interferences caused by different tissues were avoided. Consequently, the phantom contains only minimum material for the mechanical fixation of the implant (implant holder) and also has a coordinate system marked on the outside to ensure reproducible phantom placement within the scanner (Fig. 1). The implant holder is adjustable within the phantom and allows three different positions of the implant within the phantom (Fig. 1).

Implants

Representative active cardiac implants covering the whole application spectrum were selected for the measurements. The chosen implants were a cardiac loop recorder; a pacemaker; and two ICDs, one of which was optimized for reduced susceptibility artifacts (Fig.2A). Due to their functionality, the implants not only differ in size but also in the amount of ferromagnetic material.

The pacemaker and the ICDs were scanned without leads attached because prior works had demonstrated the negligible effects of the lead induced artifacts on the image interpretation[22].

Additionally, a generic setup mimicking an implant with distributed ferromagnetic material was used to investigate the impact of the distribution on size and shape of the signal void. Ferromagnetic material from ICD transformers (ferrite core material as the main ferromagnetic component) was placed in different configurations in a piece of modelling clay having the dimensions of an ICD. Two placements where investigated in more detail, while using the same amount of ferromagnetic material (Fig. 2B):

The ferromagnetic material was all in the center of the setup - configuration "C" (central);

Half of the ferromagnetic material was placed into a corner and the other half into the opposite one - configuration "D" (diagonal).

The piece of modeling clay was positioned in the phantom like a normal implant.

Measurement setup

All measurements were performed on a clinical 3T MR scanner (AchievaDS, Philips Healthcare, Best, The Netherlands). For signal reception, commercial anterior and posterior body surface coils were used (dStream Whole-Body; Philips). For the measurements, the implant was positioned at the isocenter of the scanner in the orientation 1 (Fig. 1). The pacemaker and the ICDs were positioned in the phantom with the device header cranially and pointing laterally in the direction of the Y-axis. The cardiac loop recorder was aligned on the implant holder diagonally with an anterolateral orientation. The distribution of ferromagnetic material inside the implant relative to the B0 field orientation may affect artifact extent and shape. Therefore, all measurements were performed in three orientations, each time aligning a different axis of the implant with the B0 field (Fig. 1).

In order to quantify the uncertainties of measurement regarding the signal void measurements, all implants were scanned three times with both sequences using an identical setup. Repetition of any prescans was omitted in order to avoid a potential influence of pre-scans used for optimization of imaging contrast during the clinical routine [23,24].

MR protocol and sequences

MR imaging was performed using a number of sequences covering different needs. All sequences used a Cartesian acquisition scheme in transversal orientation. An automated shimming procedure of first order was applied with the purpose of maximizing B0-homogeneity over the entire imaging volume.
A set of reference sequences was defined closely following the standard for the evaluation of MRI image artifacts as described in the ASTM publication F2119-07. This standard was established for a standardized assessment of the artifact burden [25]. The reference sequences comprised both, a multi-slice gradient echo and a multi-slice spin echo sequence with fixed values for TR, TE, receiver bandwidth, spatial resolution and flip angle. The reference gradient echo sequence deviated in its TR setting from the ASTM standard (reference sequence 50 ms, ASTM sequence 100 -500 ms) allowing for a faster image acquisition. The other deviation was a higher matrix (256 x 256) than the ASTM standard (256 x 128) for both sequences. Both factors have no influence on the artifact size of the active cardiac device. The exploration of the artifact size through the reference sequences had shown to be suboptimal. The long echo times and small bandwidth resulted in extremely large signal voids that extend beyond the edge of the phantom particularly for the larger devices 3 and 4, both of them being ICDs. Furthermore, imaging times became unsuitably long when covering the entire phantom at an isotropic resolution.

In a second set of sequences, consisting of a 3D gradient echo sequence and a multi-slice 2D turbo spin echo sequence, we performed higher resolution MR imaging with optimized imaging parameters to allow for an improved assessment of artifact size and type. Whereas a 3D approach was chosen for the gradient echo sequence, an interleaved multi-slice 2D sequence with adjacent slices (inter-slice gap = 0) was used for the TSE sequence in order to avoid an unreasonably long scan duration. Both sequence types allowed for the 3D visualization of the artifact size and shape within the boundaries of the phantom.

The effects of the implantable devices on the main magnetic field (B0) and the flip angle (B1) were assessed by mapping both characteristics with dedicated sequences. For B0 mapping, a 3D gradient echo sequence was performed twice with different echo times (delta TE = 0.1 ms). For B1 mapping, a dual TR-sequence [26] was applied (delta TR = 120 ms). The nominal flip angle was 60°. For both sequences, automatic map generation was performed on the MR system.

The dimensions of the square field of view and coverage in orthogonal (z-) direction were set to 352 mm to cover the entire cubic phantom with some additional space that was about 20% larger than the inner volume of the phantom to avoid that structures of interest extended beyond the FOV (Fig. 1).

The main imaging parameters are listed in Table 1.

Image Processing

The image data were stored in DICOM format. All subsequent processing was performed with custom software scripted in MATLAB (R2019b, Mathworks, USA). First the data were converted to 3D matrices, resampled and if needed, interpolated to obtain identically sized voxel of 2x2x2mm voxel resolution for all data sets. A brightness correction was performed by histogram stretching to compensate for differences in image brightness and to be able to apply identical gray thresholds for all data.

Evaluation of the signal void

Data segmentation

A low pass filter was applied first to eliminate the structures of the implant holder and was realized by an averaging filter with a symmetric convolution kernel of 9x9x9 voxels with all positions of the kernel filled with the value "1". Segmentation of the signal void was done by a threshold operation applied to the gray value profile of each digital image data. The histogram stretching was used to enhance the contrast. In addition, all grey values above the 95th- percentile were set to the pixel intensity of the 95th percentile. The histogram stretching did not compensate for the changes in the receive coil homogeneity.

Along the grey value profile, darker regions were encoded with values near "0" and bright regions with values near "1". The brightest signal within the volume was defined as "1" or 100%. The transition between the two extreme values has a finite and in general sigmoid shaped intensity gradient. The intensity gradient is the steepest at 50% of the sigmoid curve, thus offering the most precise cut- off between signal void and "normal" signal intensity. Accordingly, the signal void was extracted as the dark part below a threshold of 50% (see Supplemental Figure 1). To measure the actual volume of the signal void, the voxel volumes meeting the aforementioned conditions were summed up.

Metrics

To quantify the signal void two metrics were regarded: the actual volume of the signal void and the extent of the signal void's bounding box. The bounding box was determined by minimizing its volume, using the MATLAB fminsearch algorithm with six roto-translation degrees of freedom in order to calculate the volume.

B0 and B1 maps

B0 and B1 maps were analyzed by drawing circular regions of interest (ROIs; area 3cm$^2$) near the corners (17 cm from the device) and edges (12 cm from the device) of the phantom (26 ROIs per data set), as well as around the signal void (4-12 cm from the device). Measurements with a device present were compared with those obtained from a phantom without device.

**Results**

All four devices were measured with the reference sequences, the spin echo (2D-TSE) and the gradient echo (3D-TSE) sequence in all three orientations and with two repetitions.

The main influencing factors on the shape and volume of the signal void are the chosen sequence types and the implant type itself (Fig. 3). In general, gradient echo-based sequences provoke larger signal void volumes than spin echo-based sequences. Compared to the ASTM reference sequences, the optimized 3D-TFE sequence caused signal voids that were smaller by at least a factor of four for the four tested implants. These smaller signal voids do not exceed the boundaries given by the phantom size, which allows the analysis of the three-dimensional signal void shape (Fig. 4). Similarly, the optimized 2D-TSE resulted in smaller signal void volumes in comparison with the reference sequences. However, the reduction in the signal void caused by the 2D-TSE depends on the size of the signal voids in the reference sequences. Whereas small artifacts are not significantly reduced, larger artifacts in the reference sequence are reduced by a factor of two when the 2D-TSE sequence is used (Supplemental Fig. 2). Additionally, the shape of the signal void is predominantly influenced by the direction of the B0 field, whereas the orientation of the implant within the isocenter has little impact (Supplemental Fig. 3 and 4).

When the ICDs (Dev 3 and Dev 4) were measured in orientation 2, the 3D-TFE sequence caused signal voids up to 10% larger than when measured in the two other orientations (Fig. 5). The measurements with the generic setups showed that the positioning of the ferromagnetic material within the device itself influenced the artifact extent. The two generic setups showed similar signal void shapes regardless of the positioning of the ferromagnetic material. Each ferromagnetic element caused a separate but identically shaped signal void. Depending on the relative positioning of each ferromagnetic element these twin-like structures overlap but still were visually distinguished (Fig. 6). Interestingly, the total volume of signal void caused by the twins (configuration D) is less than the volume of the signal void in configuration C. This applies for both the spin echo and the gradient echo sequences. However, the bounding box dimensions for the configuration D are larger when compared to the configuration C (Fig.2).

Changes in B0 were more pronounced with larger devices. Near the corners or edges of the phantom, there was no major noticeable shift in B0 for devices 1 and 2. Closer to the devices, shifts of -580 Hz to +260 Hz and -760 to +435 Hz were measured for device 1 and 2, respectively (Fig.7).

For devices 3 and 4, there were considerable shifts in B0 beyond the visible signal void up to the edges of the phantom. In these border zones, maximum B0 differences were as much as +1160 Hz for device 3 and as much as +1440 Hz for device 4. These maximum offsets occurred for orientation 2 in z-direction, i.e., in the direction along the bore. Closer to the devices, offsets were even larger: -1280 Hz to +1420 Hz for device 3, and -1330 Hz to +1220 Hz for device 4.

Regarding the B1 measurements, none of the tested devices caused a so-called hot spot: the maximum B1 value found was 107% of the nominal flip angle in the absence of a device, and 108% in the presence of a device (Fig. 7). Overall, the B1 measurements showed very little variance for devices 1,2, and 4. Whereas there was a small trend towards smaller B1 values particularly in the front slices (farthest from the front opening, "cranial"), differences were much smaller than the standard deviation over the entire phantom and not significant. For device 3, however, there was a strong trend towards smaller B1 field in cranial direction, with the areas in the corner showing a B1 field reduced by up to 35% compared to the measurements in absence of a device.

Uncertainty assessment

Without repeating the prescans, the spin echo and the gradient echo sequences showed, when repeated, a high reproducibility of the presented results with less than 2% deviations from the mean value for the spin echo sequence and less than 4% for the gradient echo sequence. Exemplarily, for Dev 4, the signal void caused by the gradient echo ranged between 1551 mm³ and 1596 mm³ and between 674 mm³ and 686 mm³ for the spin echo sequence.

**Discussion**

The first step towards optimizing CMR image quality in the presence of active cardiac implants at 3T is a systematically performed three-dimensional analysis of the artifact extent and influencing factors. The different qualities of susceptibility induced artifacts, signal void and signal distortion, demand different settings for a systematic evaluation. Therefore, this work focuses on the signal void, using a cubic phantom with homogeneous filling and a sufficiently large volume to cover the signal void. By using a homogeneous background, it is possible to describe B1-related shading and to quantify the signal void caused solely by the presence of the device and influences of different human organ tissues are avoided.

When evaluating passive and active implants in the MRI environment, the ASTM standard has been established for clinically established field strengths [25]. However, originally designed for measurements at 1.5T, the ASTM standard proved to be unsuitable for the purpose of this work, where the effects at 3T are investigated and considerably larger artifacts occur than at 1.5T. According to the ASTM standard, the artifact itself is defined as a deviation in the signal intensity of at least 30% from a reference scan without any device being present. However, this approach requires a perfectly matched pair of reference and artifact scans without the influence of repeated prescans after repositioning. In a clinical operation mode, it is not possible to omit these prescans. Additionally, it is not possible to apply the ASTM standard definition to an in vivo application in patients with already implanted devices due to the lack of a reference scan. In contrast, the approach chosen for the presented work uses the grey value profile of each image data set, thus omitting the necessity of reference scans. The sequence protocols of the ASTM standard are set to maximize the artifact size, causing signal voids that exceed the boundaries of the phantom thus limiting the analysis of the signal void shape. The sequences optimized for our phantom, though inducing overall smaller signal voids, allowed for an image quality and spatial resolution appropriate for evaluation with high confidence at 3T. The benefit of this optimized approach is evident when focusing on devices with larger numbers of electrical components, e.g. the ICDs (Dev. 3 and 4). Additionally, the chosen sequences are paradigm sequences and can be used as foundation for other, clinically used sequences.

For both the spin echo and gradient echo sequences, the ICDs (devices 3 and 4) caused the largest artifact volumes, with device 4 causing even larger signal voids than device 3. The gradient echo sequence caused markedly larger artifacts than the spin echo sequence, and orientation 2 (X: left-right, Y: top-down, Z: foot-head) caused larger artifacts than the other orientations. The differences between Dev3 and Dev4 highlight the role of the ferromagnetic components within the active cardiac devices. The artifact sizes differ significantly whereas both devices offer similar technical features. The measurements of the generic setups demonstrate the influence of the device design on the artifact volume which is based on the distribution of the electric components within the device. These findings showed the potential that lies in the modification of the device setup with regard to the modification of the artifact burden. Although the design of the implants is not within the influence of the CMR user, information on the behavior of a specific device might help to plan the CMR examination. Additionally, it can be assumed that when a specific device is exposed to scanners with other clinical field strength, e.g. 1.5T, the shape of the signal void would remain unchanged whereas the void volume would depend on the field strength.

B0 measurements showed the expected increase in B0 shift with larger devices. It is notable that, even outside of the signal void, considerable B0 shifts may be present. This has a direct effect on any MR measurement employing a frequency-dependent pulse, such as fat suppression or an inversion pulse. This pulse type is used for instance in black blood imaging or LGE. If the preparatory pulse bandwidth is chosen too narrow, it may result in insufficient magnetization preparation and lead to image artifacts. B1 was found to be less affected by the presence of a device. Changes in the B1 field were mainly limited to the corners of the phantom and did not occur at the isocenter. However, at the border zones, the changes in the B1 field might affect the actual flip angle that consequently may be smaller than the nominal flip angle. Again, this is expected to affect features such as fat suppression or inversion pulses. The absence of B1 hot spots supports the assumption that, also in the presence of a device, no unwanted heating occurs provided that the restrictions prescribed by the MR conditional device are observed. It can be concluded that the orientation of the implant relative to the B0 field has a minor effect on the size and the shape of the signal void that in practice most likely can be neglected.

All used implants have a titanium housing. In contrast to other materials that are ferromagnetic, paramagnetic titanium produces rather markedly smaller susceptibility artefacts itself[27]. The eddy currents induced in these housings by the gradient system, however, produce additional artefacts due to the magnetic field related to these currents. These are superimposed on the susceptibility artefacts and are not separated on this study.

The outcome of the current study is still limited as we obtained data from only four implants, and measurements were performed with deactivated devices originating from the same manufacturer.

Additionally, the setup of our phantom measurements deviates from the clinical situation because a homogeneous background was used rather than an inhomogeneous background representing different organ tissue, and the chosen sequences were generic ones, from which many currently used cardiac sequences derive. However, the main conclusions are considered of general validity. The quantification of other artifact qualities, such as distortion require a different measurement set-up, and thus will be analyzed separately.

Conclusion

The presented work establishes an analysis algorithm that systematically quantifies the device related signal void. Additionally, the size and shape of the signal void obtained in this work can be helpful in estimating the relative size and shape of signal voids in vivo. The results not only highlight the effect of the sequence type on the artifact extent, but also the relevance of the device type and design on the artifact burden.

Acknowledgment

We would like to thank I. Perdijk for her support in performing the experiments.

Conflict of interest:

Dr. Weiss is an employee of BIOTRONIK SE & Co. KG, Berlin, Germany, and Dr. Weber is an employee of Philips GmbH, Hamburg, Germany. Prof. Bauer is a scientific advisor for BIOTRONIK SE & Co. KG, Berlin, Germany. Also, this work has been partially funded by BIOTRONIK SE & Co. KG, Berlin, Germany.

**Figures**

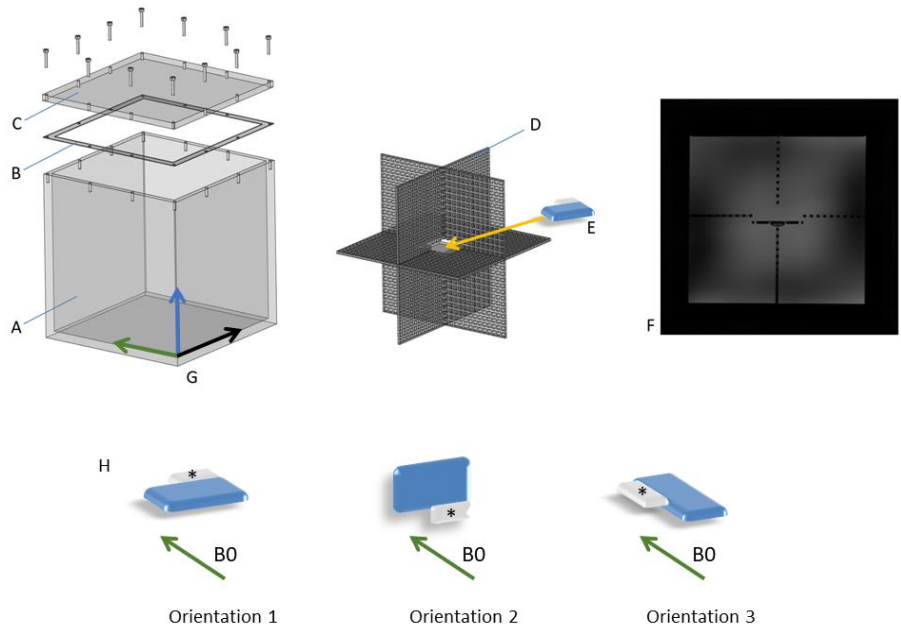

Figure 1:

The phantom consists of an acrylic canister A, closed by the lid C. The seal B ensures a water-tight closure of the phantom and prevents spills of the filling during positioning. D shows the acrylic implant holder. The implant E fits into the central space of the implant holder. F shows a scan through the phantom, showing a central section of the implant holder. The implant holder F allows three different orientations of the device relative to the B0 field (H). Changes in the orientation were achieved by turning the implant holder within the phantom. The coordinate system (G) on the phantom guided the repositioning (green arrow: Z- axis, following the direction of the B0 field, X- and Y- axis (black and blue arrows), perpendicular to the Z- axis). The asterisks mark the device header. The blue part is the implant's main body.

Orientation 1 was defined by X: left-right, Y: foot-head, Z: bottom-up. Orientation 2 was defined by X: left-right, Y: top-down, Z: foot-head. Orientation 3 was defined by X: foot-head, Y: right-left, Z: bottom-up.

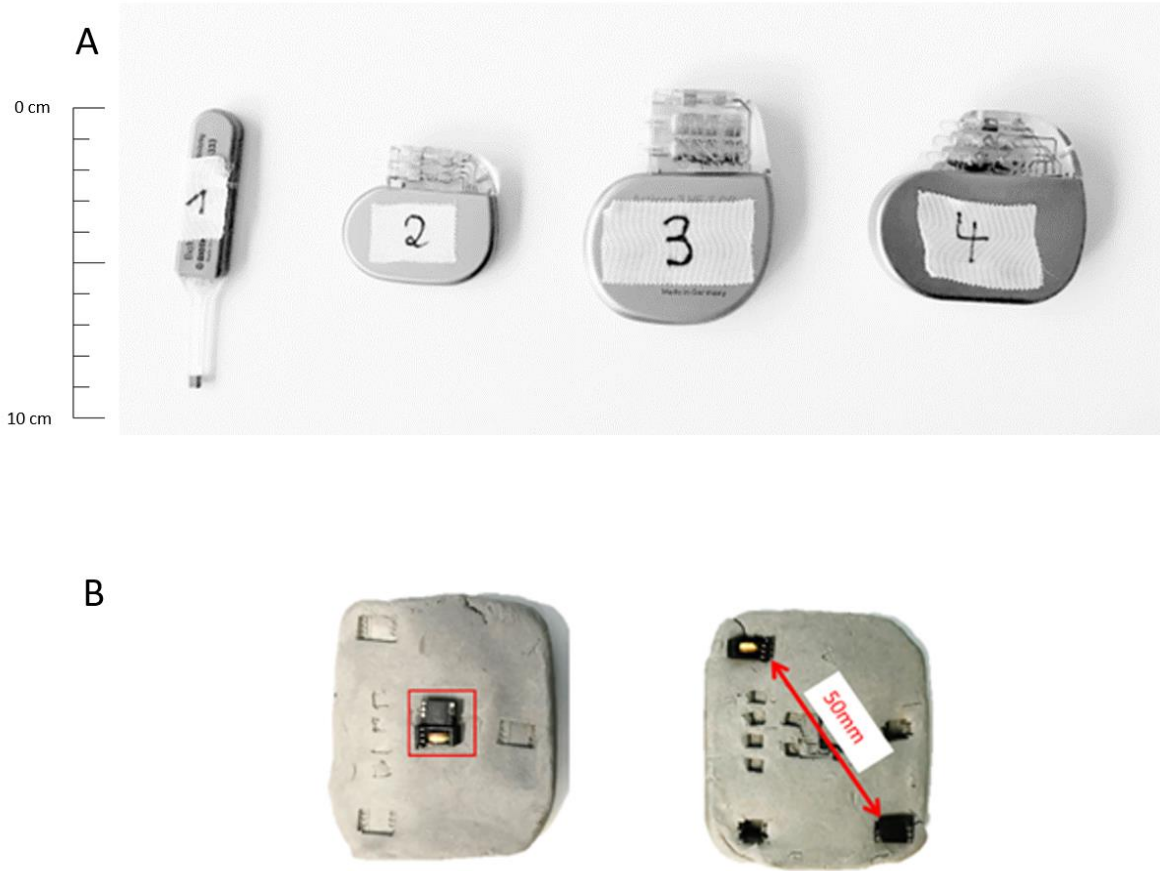

Figure 2

A) Active cardiac devices used for the measurements. All implants are manufactured by BIOTRONIK, Germany. Dev. 1: Cardiac loop recorder ("Biomonitor 2"), Dev.2: Pacemaker (Enticos 4 DR), Dev.3: ICD (Ilesto 7 HF-T), Dev.4: ICD (Activor 7 HF-T QP). The scale bar unit is cm.

B) Generic setups: The ferromagnetic material is either all placed in the center (configuration C) or with one half in one corner and the other half in to the opposite corner to be at the maximum distance possible in a state-of-the art ICD (configuration D).

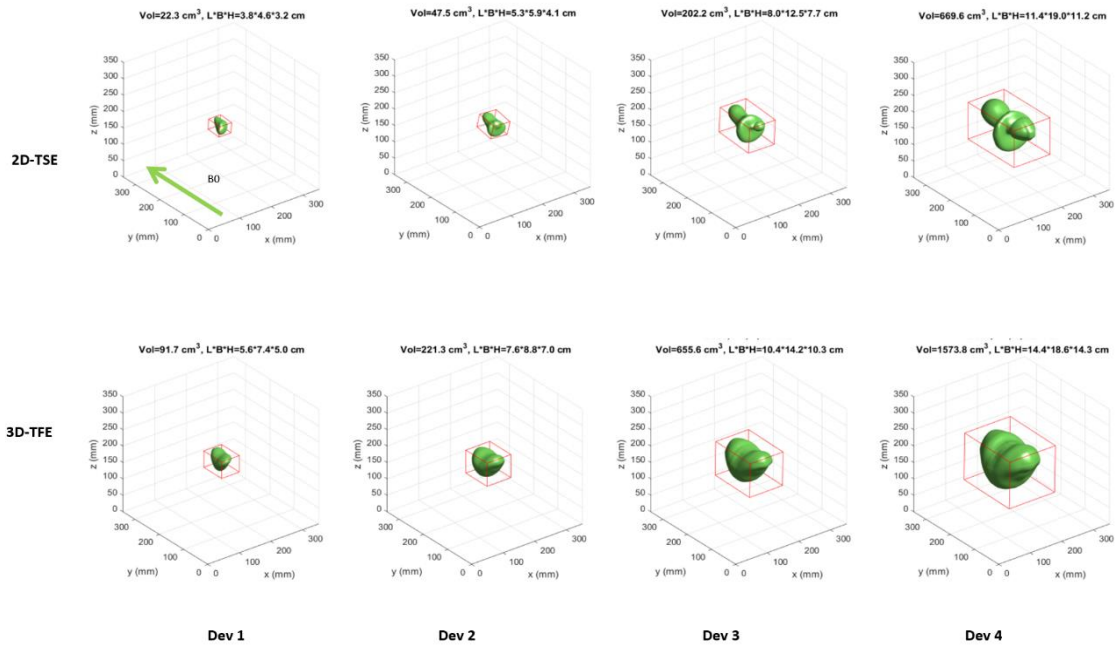

Figure 3:

Segmented signal voids induced by the implants Dev1-4 scanned with the 2-DTSE and the 3D-TFE sequences in orientation 1. The red box in each coordinate system represents the boundary box. The absolute values of the boundary box are given below the coordinate system. Dev 1: cardiac loop recorder. Dev 2: Pacemaker, Dev 3 and 4: ICDs. For results in orientation 2 and 3 see Supplemental Figure 2 and Supplemental Figure 3.

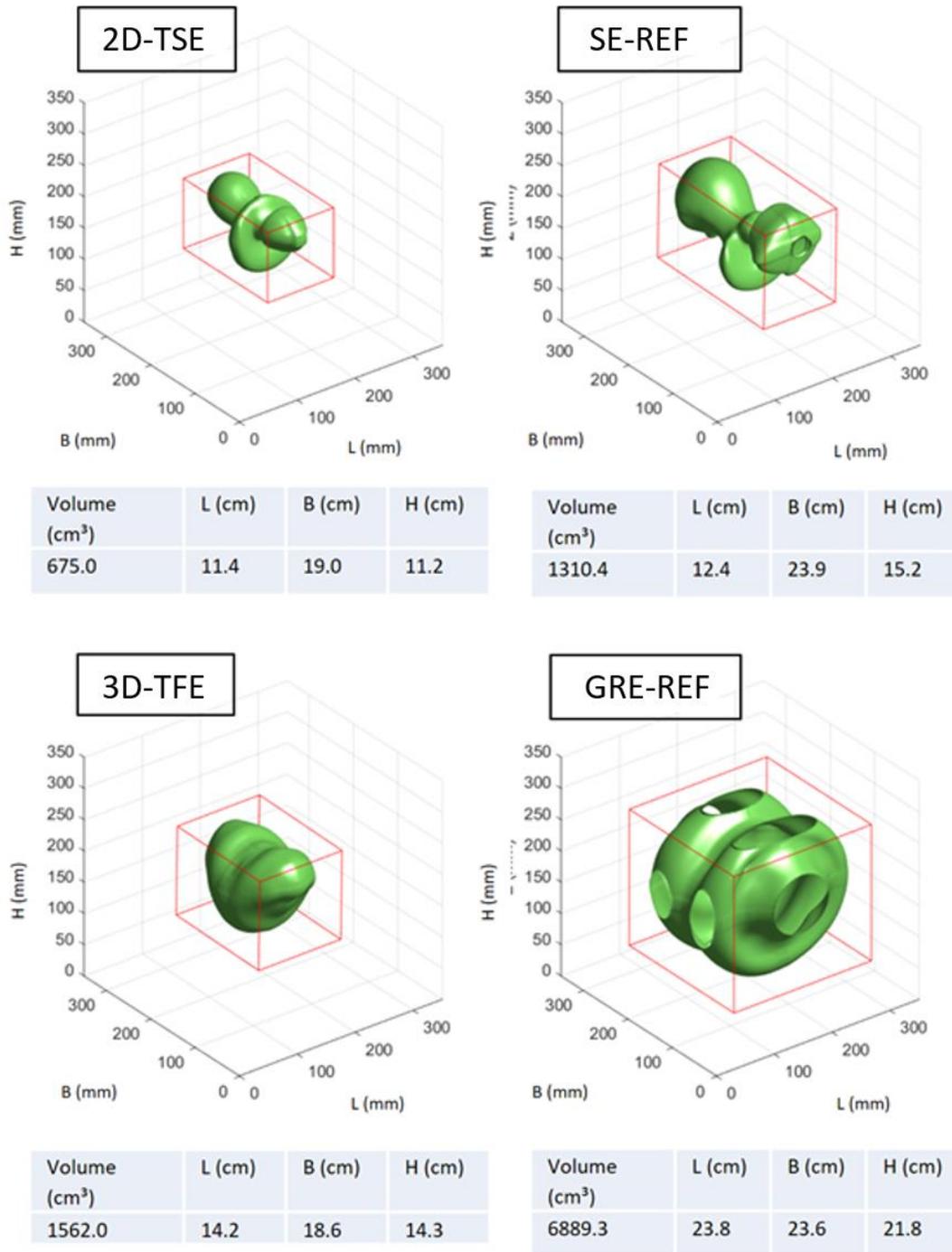

Figure 4: Morphology of the signal voids produced by the 2D-TSE and 3D-TFE sequence as compared to reference sequences, measured in orientation 1 and with Dev 4.

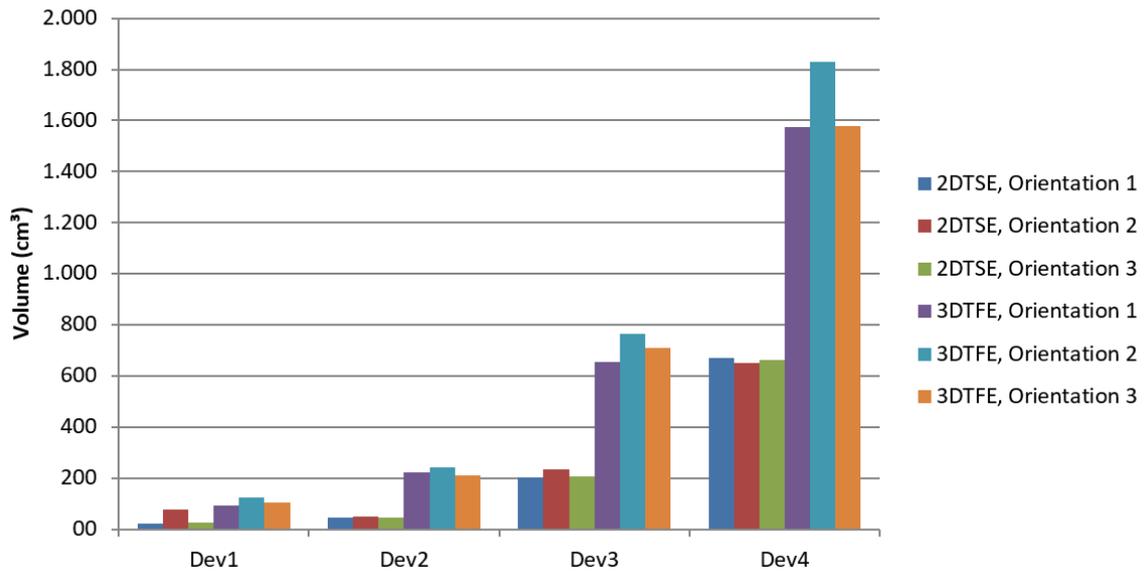

Figure 5: Comparison of signal void volumes depending on the scan sequence and the device (phantom) orientation. Dev1: cardiac loop recorder, Dev2: pacemaker, Dev3 and Dev4: ICDs.

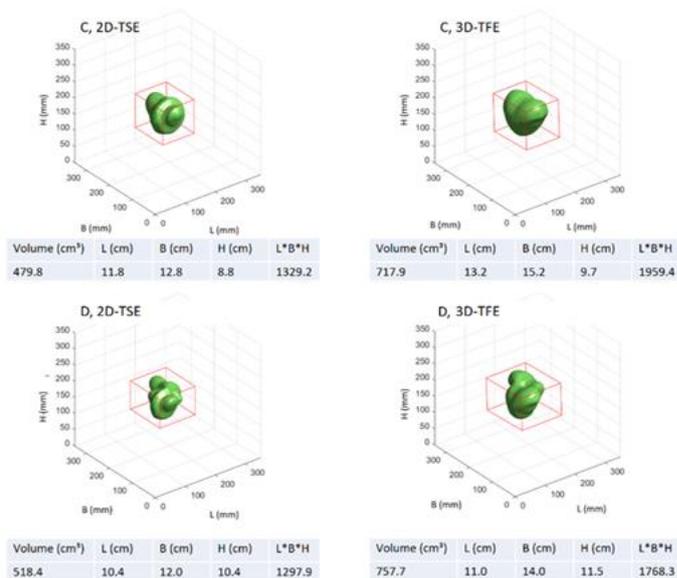

Figure 6: Generic setups and their artifacts. C describes a centric position of all ferromagnetic material, D the splitting of the ferromagnetic material into two diagonally positioned halves. All scans were performed in orientation 1. The red boxes mark the bounding box dimensions. The absolute values of the boundary box are given below the coordinate system.

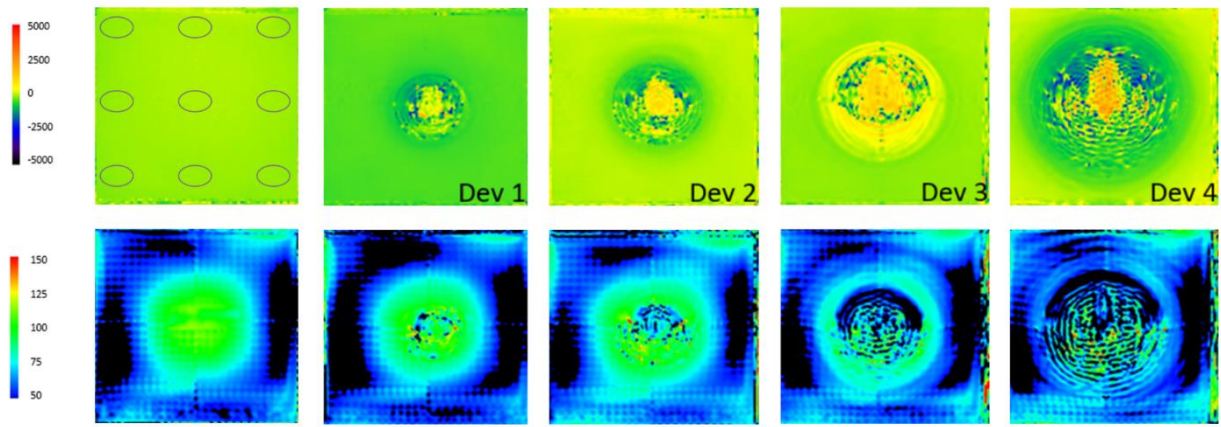

Figure 7: B0 (upper row) and B1 (lower row) maps of a central slice within the cubic phantom. The left image of each row shows the empty phantom. The scales range from -5000 Hz to +5000 Hz for B0 and from 50% to 150% relative flip angle for B1 (with 100% indicating the nominal flip angle).

The disturbances in the B1 field in the empty phantom are a standing wave effect because the edge length of the phantom is close to the ferromagnetic wave length at 3.0 T. Accordingly, disturbances in B1 in the presence of an active device are measured relatively to the empty scan.   The grey circles indicate the ROIS measured on each side of the phantom. 0 Hz is exactly at the transition from green to yellow, a bright green-yellow color thus depicts an offset close to 0 Hz. Regions with an SNR below a certain threshold are shown as black.

Table 1: Main imaging parameters

|  | Reference sequences | | Optimized sequences | | | |
|---|---|---|---|---|---|---|
|  | Grad echo | Spin echo | Grad echo | Spin echo | B0-mapping | B1-mapping |
| **Dimensions** | 2D MS | 2D MS | 3D | 2D MS | 3D | 3D |
| **Image type** | FFE | SE | FFE | TSE (5 echoes) | FFE | FFE |
| **TR (ms)** | 50 | 500 | 20 | 17560 | 30 | 30 / 150 |
| **TE (ms)** | 15 | 20 | 3.2 | 27 | 3.1 / 3.2 | 2.2 |
| **Echo spacing (ms)** | n/a | n/a | n/a | 9.1 | n/a | n/a |
| **Flip angle (deg)** | 30 | 90 | 20 | 90 | 60 | 60 |
| **Flow comp** | No | No | Yes | yes | No | No |
| **Field of view (mm)** | 352 | 352 | 352 | 352 | 352 | 352 |
| **Acq. Matrix** | 256 | 256 | 176 | 176 | 88 | 88 |
| **Nr. of slices** | 60 | 60 | 176 | 176 | 88 | 88 |
| **Spat. Resol. (acq) (mm3)** | 1.38 x 1.38 x 5 | 1.38 x 1.38 x 5 | 2 x 2 x 2 | 2 x 2 x 2 | 4 x 4 x 8 | 4 x 4 x 8 |
| **BW (Hz/pix)** | 125.2 | 125.2 | 382.4 | 473.5 | 498.4 | 498.4 |
| **Acq. Dur. (mm:ss)** | 6:28 | 17:12 | 10:24 | 21:04 | 3:52 | 11:35 |
| **Spat. Resol. (recon) (mm3)** | 1.38 x 1.38 x 5 | 1.38 x 1.38 x 5 | 1 x 1 x 1 | 1 x 1 x 2 | 2 x 2 x 4 | 2 x 2 x 4 |
| **FoV dir** | RL / AP | RL / AP | RL | RL | RL | RL |
| **Fat shift direction** | P / L | P / L | P | P | P | P |
| **Orientation** | Tra | Tra | tra | tra | Tra | Tra |
| **Max B1_rms (uT)** | 0.77 | 1.60 | 0.89 | 1.62 | 1.27 | 0.73 |

| SAR level (W/kg) | <0.3 | <1.3 | < 0.4W/kg | < 1.3 | <0.8 | <0.3 |
|---|---|---|---|---|---|---|
| Db/dt (T/s) | 47.6 | 33 | 50.4 | 38.9 | 23.4 | 50.3 |

# Supplemental File

Signal Voids of active cardiac implants at 3.0T CMR

Theresa Reiter[1,*], Ingo Weiss[2], Oliver M. Weber[3], Wolfgang R. Bauer[1]

[1] University Hospital Wuerzburg, Department of Internal Medicine I, Cardiology, Wuerzburg, Germany

[2] BIOTRONIK SE & Co. KG, Berlin, Germany

[3] Philips GmbH, Hamburg, Germany

* Corresponding author: Dr. Theresa Reiter, MD, University Hospital Wuerzburg, Department of Internal Medicine I, Cardiology, Oberduerbacher Strasse 6a, 97080 Wuerzburg, Germany. Reiter_t@ukw.de.

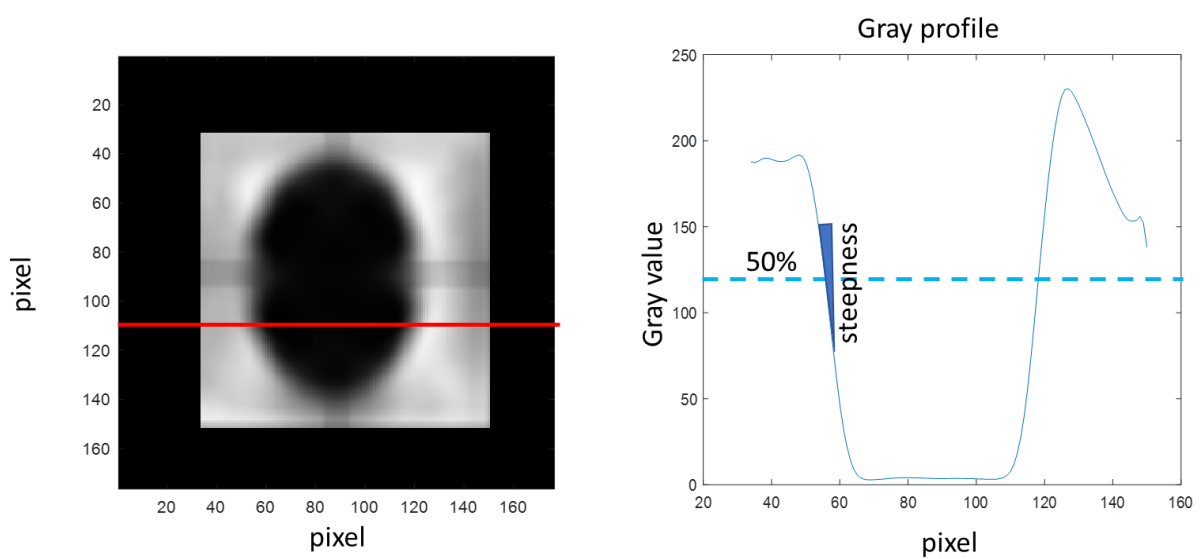

Supplemental figure 1: Example of a gray profile for a specific location. For the analysis of the signal void, the brightest signal within the whole volume data set was defined as 100%.

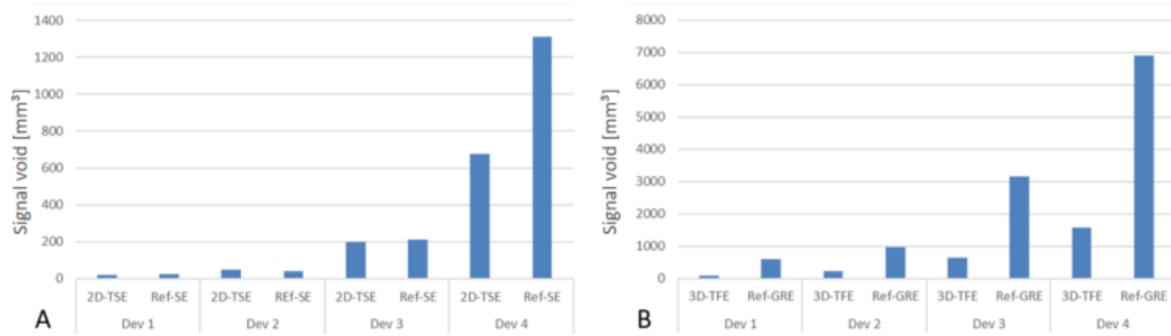

Supplemental figure 2: A) Comparison of the signal voids caused by the four devices when measured with the 2D-TSE sequence and the reference sequence. b) Comparison of the signal voids caused by the four devices when measured with the 3D-TFE sequence and the reference sequence. All measurements were performed with the orientation 1.

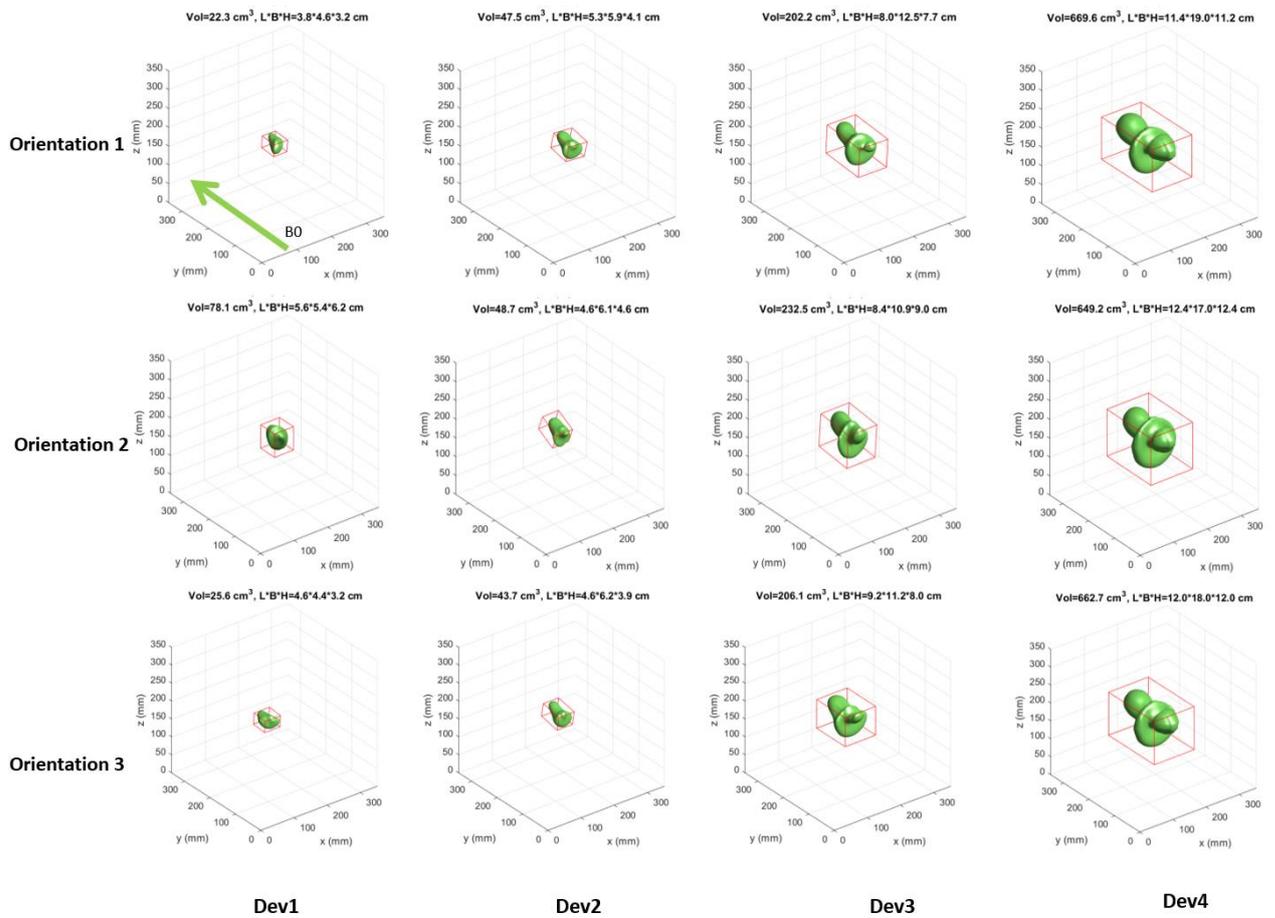

Supplemental Figure 3: 2D-TSE induced signal void volumes and bounding box for all four devices measured in three orientations.

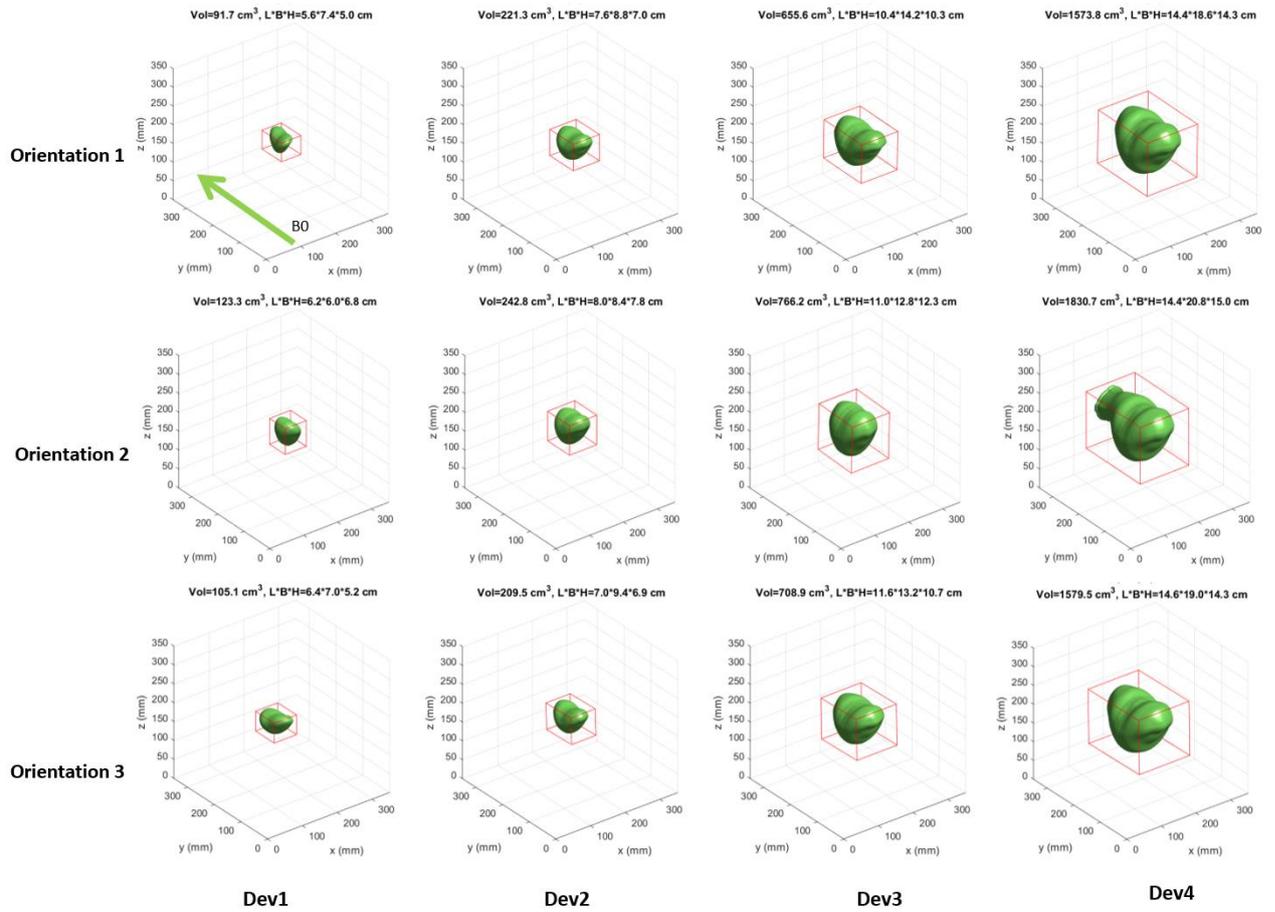

Supplemental Figure 4: 3D-TFE induced signal void volumes and bounding box for all four devices measured in three orientations.